\documentclass[12pt]{article}
\pdfoutput=1
\usepackage{latexsym}\usepackage{epsfig,amssymb,euscript}
\usepackage{amsmath} \usepackage{bbm,slashed}
\usepackage{color}
\usepackage{ulem}
\usepackage{cite}
\usepackage{graphicx}

\newcommand{\be}{\begin{equation}}
\newcommand{\ee}{\end{equation}}
\newcommand{\bea}{\begin{eqnarray}}
\newcommand{\eea}{\end{eqnarray}}

\catcode`@=12

\numberwithin{equation}{section}

\begin{document}
\begin{titlepage}

\begin{flushright}
SISSA 10/2013/FISI
%ULB-TH/09-10\\
%hep-th/yymmnnn\\
\end{flushright}
\bigskip
\def\thefootnote{\fnsymbol{footnote}}

\begin{center}
\vskip -10pt
{\LARGE
{\bf
Holographic R-symmetric \\
\vspace{0.30in}
flows and the $\tau_U$ conjecture
}
}
\end{center}

\bigskip
\begin{center}
{\large 
Matteo Bertolini,$^{1,2}$ 
Lorenzo Di Pietro$^1$  and 
Flavio Porri$^1$}

\end{center}

\renewcommand{\thefootnote}{\arabic{footnote}}

\begin{center}
\vspace{0.2cm}
$^1$ {SISSA and INFN - Sezione di Trieste\\
Via Bonomea 265; I 34136 Trieste, Italy\\}
$^2$ {International Centre for Theoretical Physics (ICTP)\\
Strada Costiera 11; I 34014 Trieste, Italy}

\vskip 5pt
{\texttt{bertmat,dipietro,fporri @sissa.it}}

\end{center}

\noindent
\begin{center} {\bf Abstract} \end{center}
\noindent
We discuss the holographic counterpart of  a recent conjecture regarding R-symmetric RG flows in four-dimensional supersymmetric field theories. In such theories, a quantity 
$\tau_U$ can be defined at the fixed points which was conjectured in \cite{Buican:2011ty} to be larger in the UV than in the IR, $\tau_U^{UV} > \tau^{IR}_U$.  We analyze this conjecture from a dual supergravity perspective: using some general properties of domain wall solutions dual to R-symmetric RG flows,  we define a bulk quantity which interpolates between the correct $\tau_U$  at the UV and IR fixed points, and study its monotonicity properties in a class of examples. We find a monotonic behavior for theories flowing to an interacting IR fixed point. For gapped theories, the monotonicity is still valid up to a finite value of the radial coordinate where the function vanishes, reflecting the gap scale of the field theory.

 \vspace{1.6 cm}
\vfill

\end{titlepage}

\setcounter{footnote}{0}

%%%%%%%%%%%%%%%%%%%%%%%%%%%%%%%%%%%%%%%%%%%%%
%%%%%%%%%%%%%%%%%%%%%%%%%%%%%%%%%%%%%%%%%%%%%%
\section{Introduction}

The renormalization group flow 
is a widely explored subject in theoretical physics, and it is object of a revived interest in the last couple of years. In a quantum field theory such flows   
interpolate, in general, between a UV fixed point describing a free or interacting CFT, and an IR fixed point which can be either free, interacting, or else a trivial one (e.g., if the theory has a mass gap).\footnote{This is a well-known theorem for two-dimensional quantum field theories   \cite{Zamolodchikov:1986gt,Polchinski:1987dy}. A full non-perturbative proof of this statement is still lacking in four dimensions, while the perturbative case is treated in \cite{Luty:2012ww}, see also \cite{Fortin:2012hn}.}  Hence, the information about the QFT in the far UV and IR is contained in the data of the conformal field theories at the two extrema of the flow, CFT$_{UV}$ and CFT$_{IR}$. An example of such data are the $c$ and $a$ central charges which appear in the one-point function of the trace of the energy-momentum tensor $\langle T^\mu{_\mu} \rangle$ when computed on a curved background. Other examples are the coefficients of two-point functions of 
conserved currents, which measure the amount of degrees of freedom charged under the corresponding symmetries.

A special role in the study of the flow is played by monotonic quantities, both as a mean to restrict or even determine the IR phases of a given theory, and to establish general properties like irreversibility. As an example, it has been proven \cite{Komargodski:2011vj} that in four dimensions the $a$-anomaly respects the inequality 
\begin{equation}
 a_{UV} > a_{IR} \label{a-theorem}~
\end{equation}
which gives support to the idea that the number of degrees of freedom get reduced in RG flows.

In the class of theories admitting a holographic description, the geometrization of the flow may help in testing inequalities such as \eqref{a-theorem} in theories with strongly  coupled fixed points, as well as to argue for stronger monotonicity theorems for interpolating functions. Indeed, on the gravity side RG flows are described by domain-wall solutions, the running quantities of the field theory being mapped to functions of the holographic coordinate. Thus, the dual gravitational background can encode, in principle, information about the {\it whole} flow. 
The particular case of the holographic a-theorem has been discussed in \cite{Alvarez:1998wr,Girardello:1998pd,Freedman:1999gp,Girardello:1999bd,Myers:2010xs,Myers:2010tj}, where a quantity interpolating between the correct $a$ charges in the UV and IR has been defined in the effective 5d supergravity theory, and proved to be monotonically decreasing as a function of the radial coordinate in the bulk. 

Our aim in this paper is to apply a similar approach to a recently proposed conjecture regarding R-symmetric RG flows of 4d supersymmetric QFTs. In such theories a quantity $\tau_U$ can be defined, which is independent of $a$, and which is related to the two-point function of a non-conserved flavor (i.e. non-R) current. This can be understood as follows. In a SCFT the superconformal R-symmetry  $\tilde R$ is the one  which, among the possibly many R-symmetries, maximizes the combination of 't\,Hooft anomalies $a(R)=3{\rm Tr}R^3 - {\rm Tr}R$ \cite{Intriligator:2003jj}. Perturbing the SCFT by a relevant deformation or a VEV of some scalar operator, there is yet another R-symmetry, $R$, which is singled out: the one that maximizes $a$ on the restricted set preserved by the flow (taking into account possible extra abelian factors coming from non-abelian symmetries broken by the relevant deformation \cite{Bertolini:2004xf}). The combination $U\equiv \frac32(R-\tilde R)$ yields a flavor symmetry which is broken 
along the flow but %asymptotically 
preserved at the superconformal fixed points. %\footnote{The R-symmetries $R$ and $\tilde R$ cannot coincide as a consequence of the a-theorem. Indeed, if they did, $\tilde R$ would be conserved all along the flow and one could use 't\,Hooft anomaly matching to prove that $a_{IR}\geq a_{UV}$. We thank Matthew Buican for a discussion on this point.} 
The associated current $U_\mu$ is a well-defined gauge invariant operator only for theories admitting a Ferrara-Zumino (FZ) multiplet \cite{Komargodski:2010rb}. For this reason, in what follows we will focus on theories in which such multiplet exists. We can thus define the coefficient $\tau_{U}$ of its two-point function in the SCFT
\begin{equation}\label{footnote}
 \langle U_\mu(x) U_\nu(0) \rangle=\frac{\tau_{U}}{(2\pi)^4}(\partial^2\eta_{\mu\nu}-\partial_\mu\partial_\nu)\frac{1}{(x)^4}~,
\end{equation}
where, due to unitarity, $\tau_U>0$. 
In \cite{Buican:2011ty} 
it was proposed that
\begin{equation}
\label{buicanconj}
 \tau_U^{UV} > \tau_U^{IR}~.
\end{equation}
This is an interesting conjecture as, if true, it would give information on the amount of emergent symmetries in the IR for a large class of QFTs. Indeed, from its very definition one sees that $\tau_U^{IR}$ can only get contribution from emergent $U(1)$ symmetries, so that the inequality \eqref{buicanconj} can be seen as a bound on the amount of them. 
More generally, $\tau_U$ is a quantity which can provide useful information about IR properties of a QFT which in some cases cannot be captured by $a$. A nice such examples was already  discussed in \cite{Buican:2011ty}, where the inequality \eqref{buicanconj} was used to determine the IR phase of the Intriligator, Seiberg, and Shenker theory \cite{Intriligator:1994rx} (see \cite{Buican:2012ec} for other interesting applications).  

In what follows, we will not attempt to prove the above conjecture, but rather see whether a holographic quantity, interpolating between $\tau_U^{UV}$ and $\tau_U^{IR}$ can be defined which is monotonically decreasing along the flow. We start in section 2 by reviewing a number of facts, well known to supergravity experts, which will enable us  to give a precise geometric characterization of R-symmetric domain walls in 5d gauged supergravity. This analysis will let us define, in section 3, a bulk quantity in terms of geometric objects of the 5d gauged supergravity (hence, ultimately, of the holographic coordinate $r$), $\tau_U(r)$. Such quantity is a natural generalization of $\tau_U$ out of the fixed points,  which correctly interpolates between $\tau_U^{UV}$ and $\tau_U^{IR}$, and whose monotonicity properties we want to study. In section 4, we will focus on the minimal gravitational set-up having all necessary ingredients to describe supersymmetric and R-symmetric QFT 
flows, and see whether our holographic $\tau_U$-function does decrease monotonically through the IR. For backgrounds corresponding to flows between interacting CFTs, $\tau_U(r)$ turns out to decrease monotonically towards the IR, reaching its minimal value at the IR fixed point. A monotonically decreasing behavior is also found for flows ending in a gapped phase, which correspond to singular  
gravitational backgrounds. Here, however, $\tau_U(r)$ becomes zero at some finite value of the holographic coordinate, well before reaching the singularity, in agreement with expectations for theories with a mass gap. In the vicinity of the singularity the monotonic behavior of $\tau_U(r)$ gets in fact reversed, but this corresponds to scales where the supergravity description is no more reliable. Section 5 contains a summary and a discussion of our results.
 
%%%%%%%%%%%%%%%%%%%%%%%%%%%%%%%%%%%%%%%%%%%%%%
%%%%%%%%%%%%%%%%%%%%%%%%%%%%%%%%%%%%%%%%%%%%%%
\section{Holographic R-symmetric RG flows}

In this section we describe the gravitational setup which is dual to RG flows in supersymmetric and R-symmetric theories. We start by recalling the correspondence between flows and domain-wall solutions, in particular when four supercharges are preserved. Then we move to an analysis of global symmetries, and we focus in particular on the implications of a conserved R-symmetry for the dual domain wall solution.

%%%%%%%%%%%%%%%%%%%%%%%%%%%%%%%%%%%%%%%%%%%%%%%%
\subsection{BPS domain walls and supersymmetric flows}

Supersymmetric AdS$_5$ stationary points are interpreted via AdS/CFT as interacting superconformal four-dimensional theories and, correspondingly, BPS domain-wall solutions
as holographic dual of four-dimensional supersymmetric RG flows. Let us briefly review how this works. 

When the four-dimensional field theory is ${\cal N}=1$ superconformal, the dual background geometry is exactly AdS$_5$, and preserves eight supercharges. In the field theory, possible deformations of the SCFT preserving $\mathcal{N}=1$ supersymmetry (i.e. four supercharges) are given by F-term deformations of the action or VEVs of the lowest component of some chiral superfield. The hypermultiplets of the $\mathcal{N}=2$ five dimensional supergravity theory, which are dual to chiral operators, account for both possibilities. A hypermultiplet contains two complex scalar fields, one of them being dual to the lowest component, and the other to the F-term component of the chiral operator ${\cal O}$, see Table \ref{bb}. Therefore, for every such scalar only one of the boundary modes can be turned-on in the solution, the one corresponding to a source deformation for the field dual to the F-term, and to a VEV deformation for the field dual to the lowest component \cite{Tachikawa:2005tq}.

%%%%%%%%%%%%%%%%%%%%%%%%%%%%%
\begin{table}
\begin{center}
\begin{tabular}{|c|c|}
\hline
% & Bulk & Boundary \\ \hline
 Hypermultiplet scalars&\hskip -15pt $\varphi_1  \underset{\overset{r\to\infty}{}}{\sim} \lambda_ O\, e^{-\Delta r} + \langle O \rangle e^{-(4-\Delta) r}$ \\ 
 &$\varphi_2  \underset{\overset{r\to\infty}{}}{\sim} \lambda_F\, e^{-(\Delta+1) r} + \langle F \rangle e^{-(3-\Delta) r}$\\ \hline
 Vector multiplet scalar&$\rho \underset{\overset{r\to\infty}{}}{\sim} \lambda_J \,e^{-2 r} r + \langle J \rangle e^{-2 r} $\\ \hline
  \end{tabular}
  \caption{The scalar field/operator correspondence. $O$ is the lowest component of the chiral operator ${\cal O}$, and it is dual to the complex scalar $\varphi_1$; $F$ is the F-component, and is dual to the complex scalar $\varphi_2$; $\mbox{dim} \,{\cal O} = \Delta$. The current ${\cal J}$ is dual to a bulk vector multiplet, its scalar component $J$ being dual to the real scalar field $\rho$; obviously, $\mbox{dim} \,{\cal J}=2$ . Finally, $\lambda_O\,,\,\lambda_F$ and $\lambda_J$ are the couplings of the Lagrangian deformations.}\label{bb}
  \end{center}
  \end{table}
%%%%%%%%%%%%%%%%%%%%%%%%%%%%%

The halving of the allowed modes at the boundary is a direct consequence of the fact that the equations of motion, which are second order equations in general, can be traded with first order equations for supersymmetric backgrounds. The solution describing a supersymmetric RG flow is therefore a BPS domain wall in the five dimensional supergravity theory, with a metric of the form
\begin{equation}\label{DWmetric}
  ds^2 =  e^{2A(r)} dx^m dx^n\eta_{mn}+ dr^2~,
\end{equation}
and one or more scalars with non-trivial $r$-dependent profiles.

Besides chiral operators, one is often interested into QFT conserved currents, which are organized in linear multiplets ${\cal J}$ containing also a real scalar operator. The latter is dual to a bulk scalar sitting in a supergravity vector multiplet, see again Table \ref{bb}. Hence, in general, one has to consider a supergravity theory including, besides the graviton multiplet, both hypermultiplets and vector multiplets.

The structure of the 5d Lagrangian, whose detailed structure can be found in \cite{Ceresole:2000jd}, can be roughly summarized as follows. The scalar fields coming from hyper and vector multiplets parametrize a sigma-model with target manifold $\mathcal{M}=\mathcal{M}_{vec}\times \mathcal{M}_{hyp}$, the vector fields gauging a subgroup of the isometries of $\mathcal{M}$. We denote the hyperscalars (in real components) by $q^X$ with $X=1,\dots 4n_H$ and the vector-scalars by $\rho^x$ with $x=1,\dots,n_V$, where $n_H$ and $n_V$ are the number of hyper and vector multiplets, respectively. The Lagrangian enjoys an $SU(2)$ R-symmetry, not to be confused with the $U(1)$ R-symmetry of the boundary theory. An important object is the $SU(2)$ triplet of {\it prepotentials} $P_I^{r}(q)$, which are functions of the hyperscalars and determine the coupling of the doublet of gravitino fields to the gauge fields
\begin{equation}
\label{gravg}
gA_\mu^I P^r_I (\sigma_r)_{\;j}^{i}\,\psi_\nu^j \subset  ({\cal D}_\mu \psi_\nu)^i ~.
 \end{equation}
The index $I = 0,1,\dots,n_V$ runs over the bulk gauge fields/the global symmetries of the boundary theory (associated to the $n_V$ vector multiplets plus the graviphoton), and $g$ is the gauge coupling. A second relevant object is a set of functions of the vector-scalars denoted by $h^I(\rho)$, in terms of which ${\cal M}_{vec}$ is defined by the equation
\begin{equation}
 C_{IJK}h^I(\rho)h^J(\rho)h^K(\rho)=1~,
\end{equation}
where $C_{IJK}$ are constant coefficients related by holography to triple 't\,Hooft anomalies of the QFT currents. The metric in the space spanned by the $h^I$'s is
\begin{equation}
 a_{IJ} \equiv -2C_{IJK} h^K+3h_Ih_J\quad ,\quad h_I\equiv C_{IJK}h^Jh^K~,
\end{equation}
which is in turn a function of the vector-scalars and determines the kinetic term of the bulk gauge fields
\begin{equation}
a_{IJ}F^{I}_{\mu\nu}F^{\mu\nu J}~ , 
\end{equation}
where $a_{IJ}$ and its inverse are used to raise and lower vector indices. Notice that $h_I h^I=1$.

The whole RG flow can be visualized as a curve on the scalar manifold $\mathcal{M}$ parametrized by the radial coordinate $r$. The BPS first-order equations for such curve reads, in terms of the whole set of scalars $\phi^\Lambda = \{ q^X, \, \rho^x \}$
\be
\label{bpsw}
\phi'^{\Lambda} = - 3g \,g^{\Lambda\Sigma} \partial_\Sigma W
\ee
where $g_{\Lambda\Sigma}$ is the metric on ${\cal M}$ and the superpotential is
\begin{equation}
 W\equiv \sqrt{\frac{2}{3}} \, | P^r | , \,  P^r\equiv h^I P_I^r~,
\end{equation}
where $|\,\cdot\, |$ denotes the norm of the three-vector in the adjoint of $SU(2)$. The fixed points of the flow on $\mathcal{M}$ are the stationary points of the superpotential, $W_{s.p.}$, and they correspond to interacting superconformal fixed points of the dual field theory.

Let us consider a flow that originates from a certain stationary point. When the whole domain of the solution is considered, either the curve ends in a different stationary point, or it goes to infinity in the scalar manifold (which is not compact).  In the former case the dual RG flow has interacting fixed points both in the UV and the IR, and in the vicinity of the stationary point(s)
\bea
 A(r) &\sim& \frac{r}{L_{UV}}\quad r\rightarrow+\infty\\
 A(r) &\sim& \frac{r}{L_{IR}}\quad r\rightarrow-\infty\,,
\eea
where $L_{UV/IR} = (gW_{s.p.})^{-1}$ are the radii of the asymptotic AdS geometries, related by holography to $a_{UV/IR}$ \cite{Henningson:1998gx,Gubser:1998vd}. In this case, the geometry can be weakly curved along the whole flow. In the second case the geometry runs into a singularity in the IR. The 5d supergravity approximation breaks down in the vicinity of the singularity, and in principle one should go beyond it to study this case. Still, these singular geometries can correctly encode many properties of QFTs which flow to a gapped phase \cite{Girardello:1999bd}. 

%%%%%%%%%%%%%%%%%%%%%%%%%%%%%%%%%%%%%%%%%%%%%%
\subsection{Symmetries and Killing vectors}

An important property of field theories are global symmetries. Some of them can be exact at all scales, while other can be preserved only at very small or very large scales. In other words, if we start with a CFT$_{UV}$ with a certain global symmetry group, only a subgroup will be preserved by the deformations/VEVs and remain exact along the whole flow. Approaching the IR some symmetries can emerge, hence the CFT$_{IR}$ can be more symmetric than the flow is, in general.

These facts can be easily read from the dual supergravity domain-wall solution. Global symmetries are mapped to gauge symmetries of the bulk theory, which act on the scalar manifold as isometries. In order for the stationary points to preserve a symmetry, they must be left invariant by the action of the isometry, and therefore the corresponding Killing vector $K^I$ must vanish at such points. The same is true for the curve on $\mathcal{M}$ representing the flow: if the curve lies on a submanifold where some of the Killing vectors vanish, then it is left invariant by the corresponding isometries, and those are the symmetries of the flow. 

For our purpose, we can and do restrict to an abelian $U(1)^{n_V+1}$ global symmetry, for which we pick a basis ($A^I$, $K_I$). In such a case,  the fields in the vector multiplets cannot be charged under an abelian symmetry, so the gauge fields will act non-trivially on the hyperscalars, only. Of course, for this to be done, the $\mathcal{M}_{hyp}$ manifold should enjoy at least a $U(1)^{n_V+1}$ isometry group. The derivatives of the $\sigma$-model are made covariant as
\bea
 \mathcal{D}_\mu q^X &=& \partial_\mu q^X + g A^I_\mu K_I^X~,
\eea
where $X=1,\dots, 4n_H$, $n_H$ being the number of hypermultiplets. From this we see that 
the scalar-dependent mass term is proportional to
\begin{equation}
g_{XY}K^X_I K^Y_J A^I_\mu A^{\mu J} \subset  g_{XY} {\cal D}_\mu q^X \,{\cal D}^\mu q^Y~,
\end{equation}
where $g_{XY}$ is the metric on ${\cal M}_{hyp}$. Since the above expression depends on the scalars, the gauge symmetry can be higgsed or exactly realized on the supergravity solution depending on the scalar profiles. 

There are some additional constraints on the quantities defining the 5d supergravity theory, which are required by the consistency of the gauging with supersymmetry, and will be useful in what follows. In particular, the Killing vectors are related to the prepotentials by
\begin{equation}
 D_X P_I^r = \mathcal{R}^r_{XY}K^Y_I~,
 \end{equation}
 where $D_X$ is the covariant derivative with respect to an $SU(2)$-connection on $\mathcal{M}_{hyp}$, whose curvature is $\mathcal{R}$. In the case of abelian symmetries that we are treating, also the following quadratic relation holds
\begin{equation}\label{Kconstraint}
 \frac{1}{2} \mathcal{R}^r_{XY} K_I^X K_J^Y=\epsilon^{rst}P_I^s P_J^t.
\end{equation}
 
%%%%%%%%%%%%%%%%%%%%%%%%%%%%%%%%%%%%%%%%%%%%%%
\subsection{Holographic characterization of the R-symmetry}

Let us now come to the characterization of an R-symmetry from the dual supergravity perspective. If the gravitino is charged under a gauged $U(1)$, then this $U(1)$ is  an R-symmetry of the field theory, because the operator dual to the gravitino is the supersymmetry current, which in turn can only be charged under an R-symmetry. From eq.(\ref{gravg}),  we then see that the condition for a symmetry $r^I K_I$ to be an R-symmetry is that $r^IP_I^r \neq 0$.

%%%%%%%%%%%%%%%%%%%%%%%%%%%%%%%%%%%%%%%%%%%%%%%%%%%%%%%%%%
\subsubsection{Superconformal fixed points and $U(1)_{\tilde{R}}$ symmetry}

On an AdS$_5$ background, where the scalars are frozen to constant values corresponding to some stationary point of the superpotential, the metric takes the form
\begin{equation}\label{AdSmetric}
 ds^2 = e^{2\frac{r}{L}} dx^m dx^n\eta_{mn}+ dr^2~,
\end{equation}
where the AdS$_5$ radius is given by $L=(gW_{s.p.})^{-1}$, $g$ being the gauge coupling. The condition of stationarity of the superpotential, due to supersymmetry, can be traded with a set of algebraic conditions \cite{Ceresole:2000jd,Ceresole:2001wi}
\begin{subequations}\label{fixedpoints}
\bea
h^I(\rho)K_I^X(q)&=&0 \label{fixedpointa}\\
P_I^r(q)&=&h_I(\rho) P^r(q,\rho)\quad,\quad P^r \equiv h^I P_I^r~.\label{fixedpointb}
\eea
\end{subequations}
The first condition \eqref{fixedpointa} ensures that a Killing vector field vanishes at the stationary point, which therefore enjoys a $U(1)$ symmetry.  
The second condition states that at the stationary point the $(n_V+1)$-vectors $P_I^r$ are all proportional to $h_I$. Via eq. \eqref{gravg} this implies that the gravitino 
is charged under such $U(1)$ symmetry, which is then an R-symmetry.  Hence, the above conditions are simply dual to the statement that any superconformal 
field theory has at least one R-symmetry, corresponding to the superconformal R-symmetry $U(1)_{\tilde{R}}$, as ensured by the superconformal algebra.

%%%%%%%%%%%%%%%%%%%%%%%%%%%%%%%%%%%%%%%%%%%%%%
\subsubsection{Consequences of an R-symmetry along the flow}\label{R-flow}

The existence of a preserved R-symmetry along the flow can be expressed by the following two conditions:
\begin{itemize}
\item{There exists a Killing vector which vanishes on the curve representing the flow: $K_R = r^I K_I \equiv 0$, so that the corresponding gauge boson $A^R = A^I r_I /r^2$ is massless;}
\item{The gravitino is charged under the corresponding $U(1)$, i.e. $r^I P_I^r \neq 0$.}
\end{itemize}
Using these two conditions, and multiplying the constraint \eqref{Kconstraint} by $r^I$ we get
\begin{equation}
\label{Rsurf}
\epsilon^{rst} (r^I P_I)^s P_J^t = 0\,.
\end{equation}
This equation implies that along R-symmetric flows the matrix $P_I^r$ is rank one, as it is at  superconformal points, so that we can write it in the form
\begin{equation}\label{H}
P_I^r (q) = H_I (q, \rho) {\cal P}^r(q, \rho) \quad , \quad a^{IJ}H_I H_J = 1~.
\end{equation}
This factorization condition defines two functions $H_I(q, \rho)$ and ${\cal P}^r(q,\rho)$ (we have normalized $H_I$ to 1 in order for the definition to be unambiguous). Notice that while $h_I$ only depends on the vector-scalars $\rho$, $H_I$ is also function of the hyperscalars $q$. Of course, $H_I \to h_I$ and ${\cal P}^r \to P^r$ when approaching a stationary point. The distinction between an R-symmetry $r^I K_I$ and a flavor (i.e. non-R) one $f^I K_I$ can be thus rephrased as
\begin{equation}\label{R-flavor}
r^I H_I \neq 0, \, \quad f^I H_I = 0.
\end{equation}
In the following, we normalize $r^I$ by requiring the corresponding R-symmetry to give unitary charge to the gravitino, that is $g |r^I P^r_I | = 1$.

%%%%%%%%%%%%%%%%%%%%%%%%%%%%%%%%%%%%%%%%%%%%%%
%%%%%%%%%%%%%%%%%%%%%%%%%%%%%%%%%%%%%%%%%%%%%%
\section{A holographic interpolating $\tau_U$-function}

We have assembled all elements we need to construct a quantity which can extrapolate the coefficient $\tau_U$ out of the fixed points in the bulk. As a first step, if there are many R-symmetries preserved by the deformations/VEVs which turn on the flow, we must select one using the supergravity analogue of $a$-maximization in the subset of preserved R-symmetries. In theories with a holographic dual, $\mbox{Tr} R=0$ at leading order in $1/N$ \cite{Henningson:1998gx}, and the $a$ coefficient is just proportional to the $U(1)_R$ cubic 't\,Hooft anomaly. Therefore, the corresponding supergravity procedure consists in maximizing the coefficient of the trilinear Chern-Simons coupling, which is indeed dual to the cubic 't\,Hooft anomaly. In the $(K_I,\, A^I)$ basis this is simply given by the coefficients $C_{IJK}$ which define ${\cal M}_{vec}$. 

Once the preferred conserved R-symmetry has been selected, which in the following will be simply indicated by 
\begin{equation}
(K_R,\, A^R)= (r^I K_I, \, r_I A^I / r^2)\,,
\end{equation}
we still have to identify the (broken) flavor symmetry $U$ whose two-point function defines the quantity $\tau_U$. Recall that $U = 3/2 (R - \tilde{R})$ at the fixed points, where $\tilde{R}$ is the superconformal R-symmetry. The direction  of the superconformal R-symmetry in the $n_V+1$-dimensional space, after proper normalization, is given by 
\begin{equation}
\dfrac{h^I}{g |P^r|}  |_{s.p.},
\end{equation}
with $|P^r|$ denoting the norm of the 3-vector. Hence, defining the combination corresponding to $U$ as $K_U = u^I K_I$, when approaching stationary points
\begin{equation}\label{uLimit}
u^I \to \frac{3}{2}\left( r^I - \frac{h^I}{ g |P^r|} \right)|_{s.p.} \,.
\end{equation}
As explained in the previous section, the condition that the flow is R-symmetric leads us to define natural quantities $(H_I,\,\mathcal{P}^r)$ which coincide with $(h_I,\,P^r)$ at stationary points. Therefore, a natural guess for $u^I$ outside the stationary points is 
\begin{equation}\label{uI}
u^I = \frac{3}{2}\left( r^I - \frac{H^I}{g |\mathcal{P}^r|} \right)\,.
\end{equation}
Notice that $u^I H_I = 0$, as expected for a flavor symmetry. Moreover, if the preserved R-symmetry flows to the superconformal one at the IR fixed point, then by definition $u^I |_{IR} = 0$. As we will now explain, the vector $u^I$ contains all information we need in order to define an extrapolation of the quantity $\tau_U$ along the flow. Note that here we choose to determine $u^I$ by a simple inspection of its properties in the AdS limit. A more systematic approach \cite{noi} could be pursued by studying the gravity multiplet dual to the field theory R-multiplet \cite{Gates:1983nr} (see also \cite{Komargodski:2010rb}).

At a fixed point one can read the two-point functions of conserved currents from the gauge kinetic function of the dual gauge bosons in supergravity, the precise 
formula being \cite{Freedman:1998tz}
\begin{equation}
<J_{I\mu} J_{J\nu}> = \dfrac{\tau_{IJ}}{(2\pi)^4} (\partial^2 \eta_{\mu\nu} - \partial_\mu\partial_\nu)\dfrac{1}{x^4}\, , \quad \tau_{IJ} = \dfrac{8 \pi^2 L}{\kappa_5^2} a_{IJ}\,,
\end{equation} 
where $L$ is the AdS radius and $\kappa_5$ is the gravitational coupling constant%\footnote{The gauge kinetic term in the 5D supergravity action is
%\begin{equation}
%-\frac{1}{4\kappa_5^2} \int d^5x a_{IJ} F^I F^J\,.
%\end{equation}
%Therefore $\kappa_5^{-2} a_{IJ}$ has the dimension of an inverse length, and correctly $\tau_{IJ}$ is dimensionless.}
. It follows that at fixed points the $\tau$ coefficient of the generic current associated with the symmetry $v^I K_I$ is simply obtained by taking the scalar product
\begin{equation}
\dfrac{8 \pi^2 L}{\kappa_5^2} a_{IJ} v^I v^J\,.
\end{equation}
A natural extrapolation for the quantity $\tau_U$ along the flow is hence given by
\begin{equation}
\label{htuconj}
\tau_U(r) = \dfrac{8 \pi^2 L(r)}{\kappa_5^2} a_{IJ}(r) u^I(r) u^J(r)~,
\end{equation}
where $L(r) = [A'(r)]^{-1}$ interpolates between the radii of the AdS stationary points $L_{UV}$ and $L_{IR}$ and is monotonic along RG flows \cite{Alvarez:1998wr,Girardello:1998pd,Freedman:1999gp,Girardello:1999bd,Myers:2010xs,Myers:2010tj}. Notice that the metric $a_{IJ}$ appears in the supergravity action as a function of the scalars, and thus naturally inherits the $r$-dependence from the non-trivial profile of the scalars in the domain wall solution. Notice that, by construction, $\tau_U(r)$ reduces to expected the field theory value at the UV and IR fixed points.

The quantity $\tau_U(r)$ defined above is the one whose monotonicity properties we want to study. 
We did not succeed in establishing the monotonicity %of $\tau_U(r)$ 
by just looking at the BPS equations, because it turns out that its behavior does not depend only on local data,  but some additional information about the global features of the flows are needed (see below). Hence, below we will test in some concrete examples whether the quantity (\ref{htuconj}) provides a monotonically decreasing function on R-symmetric RG flows.

%%%%%%%%%%%%%%%%%%%%%%%%%%%%%%%%%%%%%%%%%%%%%%%%
%%%%%%%%%%%%%%%%%%%%%%%%%%%%%%%%%%%%%%%%%%%%%%%%
\section{Testing monotonicity: $\mathbf{U(1)\times U(1)}$ model}

In what follows, we want to consider the minimal set-up which can encode all necessary ingredients to describe non-trivial supersymmetric and R-symmetric flows. These correspond to QFTs starting from a UV fixed point with a $U(1)_{\tilde R}\times U(1)$ symmetry which is broken to a $U(1)_R$ symmetry along the flow (when this is the only IR symmetry, the theory does not have emergent symmetries: the IR superconformal R-symmetry coincides with the $U(1)_R$ preserved along the flow, and $\tau_U^{IR}=0$). 

A two-parameter family of supergravity theories admitting domain-wall solutions dual to this kind of flows exists \cite{Ceresole:2001wi}. These are $\mathcal{N}=2$ gauged supergravity models coupled to one vector multiplet and one hypermultiplet with a $U(1)\times U(1)$ gauging. The two parameters control the gauge group embedding inside the isometry group of the scalar manifold. In a sub-region of the parameter space, the theory admits two different AdS stationary points and smooth domain-wall solutions connecting them.  These solutions correspond to QFTs described by interacting SCFT in the IR. 
The model admits also other solutions, which emanate from the same UV AdS stationary point but then run to infinity in field space. These solutions, which are singular in the deep IR,  correspond to QFTs enjoying  a mass gap. We will consider both types of solutions in turn.\footnote{This same model, in a different region of parameter space, admits also other singular (but non supersymmetric) domain-wall solutions, considered in \cite{Argurio:2012cd} in the context of holographic models of gauge mediation, see also \cite{Argurio:2012bi}.}  To our purposes, a common property that any solution should have, is that the hypermultiplet runs.  This is because the FZ-multiplet, which, as already emphasized, should be present for the operator $U_\mu$ to be well-defined, is described in terms of a couple of ${\cal N}=1$ superfields, $({\cal J}_\mu, X)$, the former being a real vector superfield which contains the supercurrent and the energy-momentum tensor, the latter being a chiral superfield containing trace 
operators. The superfield $X$ vanishes at fixed points, but should be non-trivial anywhere else; in particular, it runs along an RG flow. Hence, the corresponding dual supergravity field, which is the hypermultiplet, should have a non-trivial profile on solutions describing RG flows of QFTs with well-defined operator $U_\mu$. 

%%%%%%%%%%%%%%%%%%%%%%%%%%%%%%%%%%%%%
\subsection{The model}
Let us first summarize the basic geometric ingredients of the model at hand. For a complete presentation we refer to \cite{Ceresole:2001wi}, to whose notation we attain. The scalar  manifold is
\begin{equation}\label{manifold}
\mathcal M_{vec} = O(1,1)~,~ \mathcal M_{hyp}=\frac{SU(2,1)}{SU(2)\times U(1)}~.
\end{equation}
The manifold $\mathcal M_{vec}$ can be parametrized by a real scalar $\rho>0$ and the hyper manifold by $q^X=\{V,\,\sigma,\,\theta,\,\tau\}$, for $V>0$. Since the scalar manifold is a symmetric space we can choose, without loss of generality, the first stationary point to be at
\begin{equation}\label{O1}
O_1 : \qquad q^X=(1,\,0,\,0,\,0),\quad       \rho=1\,.
\end{equation}
The theory contains two vectors: one in the gravity multiplet (the graviphoton) and one in the vector multiplet. Thus one can gauge at most a $U(1)\times U(1)$ subgroup of the $SU(2,1)$ isometries of $\mathcal M_{hyp}$. Requiring the gauged isometries $K_I$ (now $I=0,1$) to leave $O_1$ invariant means that they should belong to the isotropy group of that point, which is $SU(2)\times U(1)$. Moreover, in order for $O_1$ to be a stationary point, the $K_I$'s should be such that their prepotentials satisfy eq.\eqref{fixedpointb}. Taking into account these constraints, the residual freedom in choosing the gauging can be parametrized by two real variables ($\beta,\,\gamma$). Following 
\cite{Ceresole:2001wi} we take
\begin{align}
 &K_0 = \sqrt2 \left(k_3 + \frac{2 \gamma}{\sqrt3} k_4\right)~,
\\ &K_1 = 2 \left(  k_3 + \frac{\beta}{\sqrt3} k_4\right)~,
\end{align}
where $k_3$ is one of the $SU(2)$ generators and $k_4$ generates the $U(1)$ factor. The stationary point $O_1$ will be our UV fixed point.
Once the gauging is fixed one finds a second stationary point at
\begin{align}
O_2: \qquad q^X&=(1-\xi_*^2,\,0,\,\xi_*\cos(\varphi),\,\xi_*\sin(\varphi))\,,\quad\rho=(2\zeta)^{\frac{1}{6}}\nonumber\\ 
			\mbox{where} \quad \xi_*&=\sqrt{\frac{2-4\zeta}{3\beta-1-4\zeta}}\,,\quad \varphi\in \left[0,\,2\pi\right]\,,\quad\zeta=\frac{1-\beta}{2\gamma-1}\,.
\end{align}
More precisely, this is in fact a {\it circle} of stationary points parametrized by the value of $\varphi$. Notice that in order for $O_2$ to belong to the manifold, the critical value of $\xi$ must be smaller than $1$ (and the radicand must be positive). Requiring the existence of $O_2$ restricts the parameters to the region
\begin{equation}\label{existence}
(\beta-1)(1-2\zeta)>0\quad \cap\quad \zeta>0~.
\end{equation}
The only combination of isometries vanishing at the second stationary point is given by\footnote{Notice that $g=(L_{UV})^{-1}$ since $W$ evaluated at the UV fixed point is equal to one.}
\begin{equation}\label{Rir}
 \left. \frac{h^I K_I}{|gP^r|}\right|_{\rm O_2}=\frac{L_{UV}}{1+\zeta}\left(\sqrt2\zeta K_0 + K_1 \right)~,
\end{equation}
and should be understood as the dual of the superconformal R-symmetry of the SCFT$_{IR}$. This implies that flows ending at $O_2$ are dual to QFTs without emergent symmetries and hence with $\tau^{IR}_U=0$.

%%%%%%%%%%%%%%%%%%%%%%%%%%%%%%%%%%%%%%%%%%%%%%%%%%%%
\subsection{Flow solutions}

Supersymmetric flows starting from $O_1$ can end in a point $O_2$ of the circle of stationary points, or to infinity in the scalar manifold. Both types of solutions can be found solving the system of BPS equations defined by the superpotential of the theory, eq.\,\eqref{bpsw}. One finds that the solutions lie on the plane ($\rho,\,\xi$) defined by
\begin{subequations}\label{plane}
\begin{align}
 V&=1-\xi^2~,
\\ \sigma&=0~,
\\ \theta+i\tau&=\xi \, e^{i\varphi}~,
\end{align}
\end{subequations}
with fixed but arbitrary value for $\varphi$, which then parametrizes an exactly marginal deformation of the entire flow (including the IR fixed point). 

\begin{figure}
\center
\includegraphics[scale=0.8]{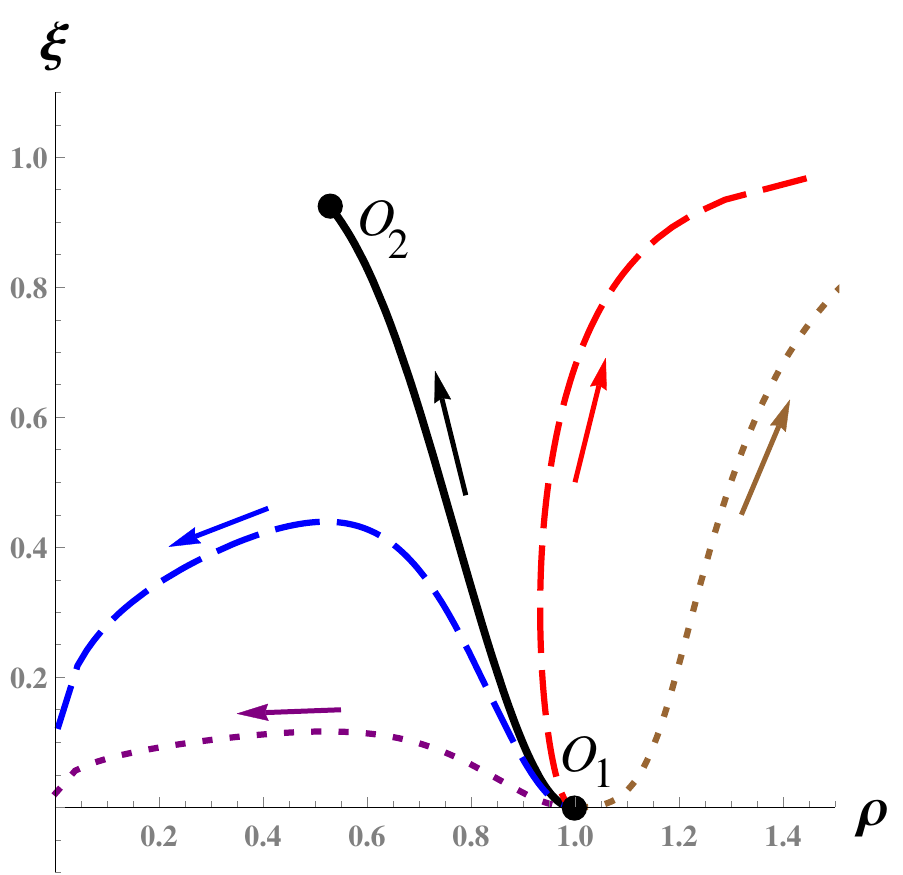}
\caption{Flow solutions for $(\beta, \zeta) =(1.1, 0.01) $. The solid black line is the only flow interpolating between the stationary points $O_1$ and $O_2$. Other flows emanating from the UV point $O_1$ are shown in dashed lines (red, blue, purple, brown), and they are singular in the IR. %The equations also admit two flows emanating from the IR stationary point $O_2$ (orange and cyan dot-dashed lines), both ending at a singularity in the IR.
}\label{flows}
\end{figure}

The system of first order differential equations for the remaining two fields can be solved numerically for any given $(\beta,\zeta)$. The curves in the ($\rho,\,\xi$)-plane corresponding to various types of flows are shown in Fig.\,\ref{flows} for a particular choice of $(\beta,\zeta)$ (other values of $(\beta,\zeta)$ provide analogous diagrams).
%%%%%%%%%%%%%%%%%%%%%%%%%%%%%%%%%%%%%%%%%%%%
%%%%%%%%%%%%%%%%%%%%%%%%%%%%%%%%%%%%%%%%%%%%
One finds that the combination of Killing vectors \eqref{Rir} vanishes on the entire plane \eqref{plane}. We can then define 
\begin{equation}\label{tauIR=0}
r^I =  \left. \frac{h^I}{|gP^r|}\right|_{\rm O_2} = \frac{L_{UV}}{1+\zeta} \left(\sqrt2 \zeta,\, 1 \right)
\end{equation}
where the normalization is such that $g^2|r^I P_I|^2=1$. According to the discussion in section \ref{R-flow} this means that the symmetry gauged by the massless vector $A^R = r_I A^I / r^2$ is dual to an R-symmetry which remains unbroken along the flow. Since no other symmetry is preserved along the flow, there is no need to perform $a$-maximization in the set of conserved R-symmetries to determine the relevant one. One can now compute the quantity $H_I,\,{\mathcal P ^r}$ as in eq. \eqref{H} and then use the definition \eqref{htuconj} to find $\tau_U(r)$. 

Before presenting our numerical results, two comments are in order. First, comparing eq. (\ref{tauIR=0}) with eq. (\ref{uLimit}) it is clear that, as anticipated, for all flows $\tau_{IR} = 0$. Therefore, the inequality (\ref{buicanconj}) should, and in fact does trivially hold. However, what we are really interested in is to test the stronger hypothesis of a monotonic behavior of $\tau_U$ along the flow. This will be our concern in the following. As a second comment, let us notice that besides flows of the types depicted in Fig.\,\ref{flows}, the theory admits few additional R-symmetric flow solutions, starting either from $O_1$ or $O_2$, characterized by the fact that there is no hyperscalar running along the flow (in fact, any flow starting from $O_2$ must fall in this class, since at that point the hypermultiplet has been eaten by the broken vector field to form a massive vector multiplet). As explained above, for flows of this kind, as far as the supergravity solution can say, the dual QFTs do not have 
a FZ-multiplet. For such theories the conjecture \eqref{buicanconj} is not expected to hold; in fact, it cannot even be formulated. Hence, we will not consider these other flows in the following. 

\begin{figure}
\center
\includegraphics[scale=0.9]{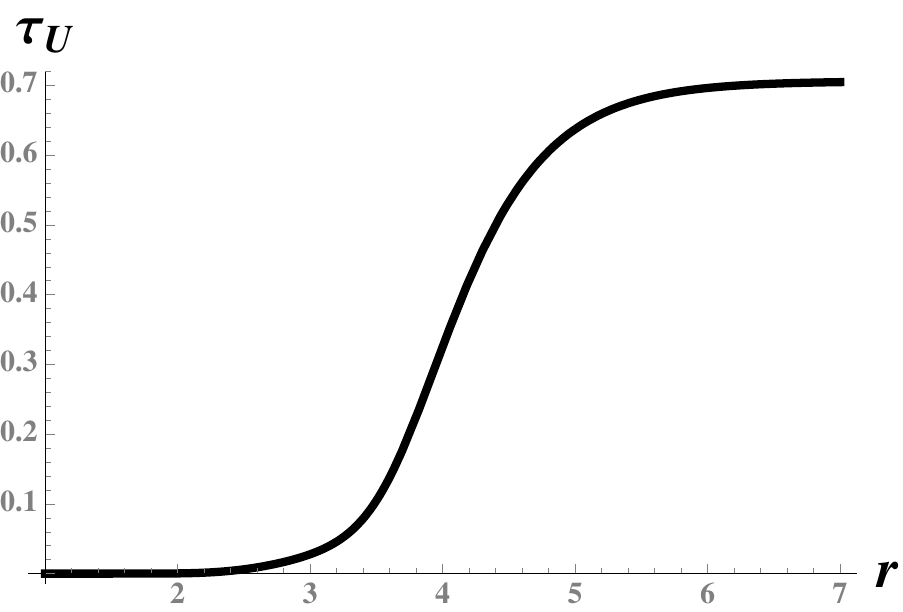}
\caption{$\tau_U(r)$ on the domain-wall solution with $(\beta,\zeta)= (1.1,0.01)$, displaying a monotonically decreasing behavior toward the IR.}\label{tauregular}
\end{figure}

\paragraph{Domain-wall solutions:}
In Fig.\,\ref{tauregular} we show a plot of $\tau_U$ evaluated on the AdS-to-AdS domain-wall solution (corresponding to the black solid line in Fig.\,\ref{flows}). 
%%%%%%%%%%%%%%%%%%%%%%%%%%%%%%%%%%%%%%%%%%
%%%%%%%%%%%%%%%%%%%%%%%%%%%%%%%%%%%%%%%%%%
The plot corresponds to a solution with specific values of $(\beta, \zeta) $. However, one can check that the qualitative behavior displaced in Fig.\,\ref{tauregular} does not change by changing the parameters ($\beta,\,\zeta$). Hence, we can conclude that on the entire class of flows interpolating between interacting fixed points, the holographic $\tau_U(r)$ function monotonically decreases toward the IR. Let us notice that, as shown in \cite{Ceresole:2001wi}, for a specific choice of $(\beta,\zeta)$ one retrieves the FGPW flow \cite{Freedman:1999gp}, which was originally obtained within ${\cal N}=8$ supergravity and whose field theory dual is the Leigh-Strassler flow \cite{Leigh:1995ep}.

\begin{figure}
\center
\includegraphics[scale=0.86]{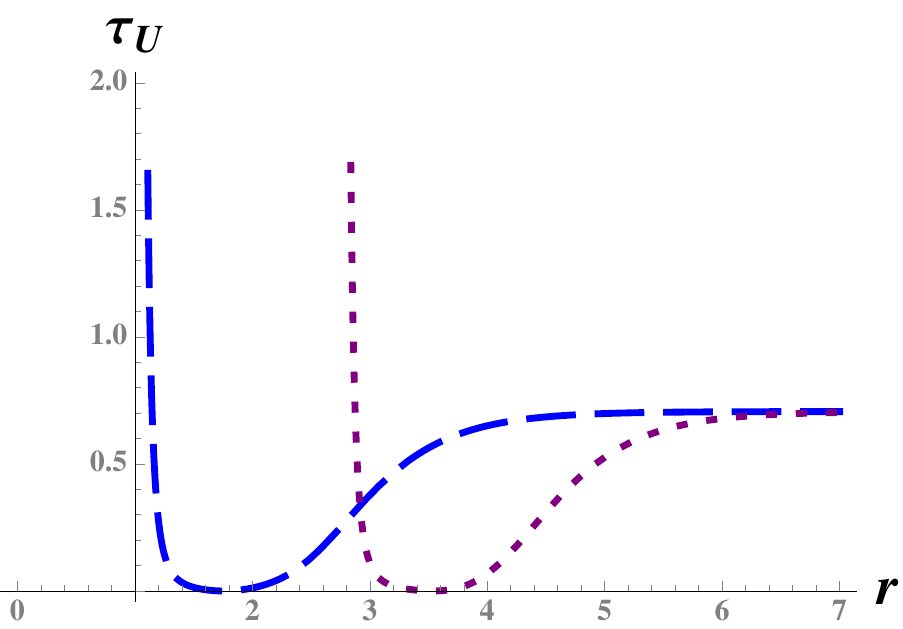}
\includegraphics[scale=0.86]{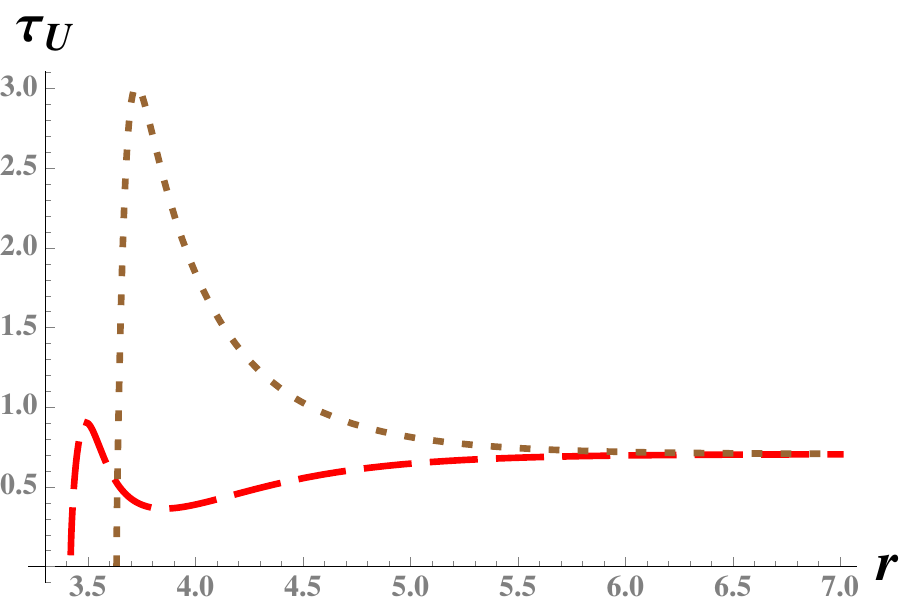}
\caption{$\tau_U(r)$ on the singular solutions for $(\beta, \zeta) =(1.1, 0.01) $. Colors and dashing refer to those of Fig. \ref{flows}. On the left the flows ending at $\rho \to 0$ ($\mathcal{V} \to - \infty$), on the right those ending at $\xi \to 1$ ($\mathcal{V} \to + \infty)$.}\label{tausingular}
\end{figure}

%%%%%%%%%%%%%%%%%%%%%%%%%%%%%%%%%%%%%%%%%%%%%%%%%%%%
\paragraph{Singular flows:}
In Fig.\,\ref{tausingular} we plot the behavior of $\tau_U(r)$ along the singular flows showed in Fig.\;\ref{flows}. A first class of flows (left plots) ends at $\rho \to 0$ in the scalar manifold (recall that by definition $\rho > 0$  so that $\rho = 0$ is not a point of the scalar manifold but is in fact at infinite distance from any point thereof). A second class of flows (right plots) ends instead at $\xi \to 1$ (the same consideration as before applies, since $\xi$ is defined to be strictly less than 1). Along flows belonging to the first class $\tau_U(r)$ starts decreasing and does so until a finite value of $r$ where it actually vanishes. In the second class there are instead flows along which $\tau_U(r)$ starts increasing from the outset, or inverts its monotonic behavior before vanishing. Another difference between the two classes of flows is the behavior of the scalar potential $\mathcal{V}$ while approaching the singularity: $\mathcal{V}\to -\infty$ in the first class, and $\mathcal{V} \to +\infty$ 
in the second. According to the  criterion proposed in \cite{Gubser:2000nd}, only the first class of singular solutions is acceptable (i.e. they are believed to admit a shielding by a horizon and/or a resolution in string theory). Hence, only flows described by the left plots of Fig.\;\ref{tausingular} are expected to have a meaning in the dual field theory and should be considered. 

We postpone to the next and last section a discussion of all above results

%In the next and last section we will discuss the above results.

%%%%%%%%%%%%%%%%%%%%%%%%%%%%%%%%%%%%%

%%%%%%%%%%%%%%%%%%%%%%%%%%%%%%%%%%%%%%%%%%%%%%%%
%%%%%%%%%%%%%%%%%%%%%%%%%%%%%%%%%%%%%%%%%%%%%%%%
\section{Summary and discussion of results}

We have considered the holographic counterpart of a field theory conjecture regarding supersymmetric and R-symmetric flows of 4d QFTs. Our goal was to define a holographic quantity $\tau_U(r)$ interpolating between the UV and IR values of $\tau_U$ and check its monotonicity properties along the flow.

We have tested our proposal on a simple 5d gauged supergravity model. Regardless its simplicity, there are a number of lessons we can learn from this minimal set-up which might have a wider validity, and hold for many supersymmetric and R-symmetric QFT flows admitting a holographic description. 

Our results show that $\tau_U(r)$ decreases monotonically towards the IR for flows interpolating between AdS critical points, corresponding to QFTs being described by an interacting SCFT in the IR. This suggests that the stronger version of the conjecture might hold in general, for this class of flows. On the other hand, singular flows, corresponding to QFTs enjoying a gapped phase, have a holographic $\tau_U(r)$ which starts decreasing towards the IR, reaches a vanishing value at some finite distance (corresponding to a finite energy of the dual field theory), and then starts increasing. Naively, one then might conclude that for QFTs not ending at an interacting fixed point the stronger version of the conjecture \eqref{buicanconj} does not hold. However, we suggest a different interpretation.

For theories enjoying a mass gap, $\tau_U$ becomes zero at a scale $\Lambda$ of the order of the masses of the lightest excitations, below which the theory is empty. Our findings simply reflect such behavior, provided we interpret the finite value of $r^*$ at which $\tau_U=0$ as corresponding to the scale $\Lambda$ of the dual field theory. This identification is supported by the fact that at radial distance $r \sim r^*$ the warp factor develops the logarithmic behavior typical of gapped phases. Below such scale, the change in the behavior of $\tau_U(r)$ as $r$ is further decreased is just an artifact of the supergravity approximation, which, by standard AdS/CFT arguments, could not anyhow be trusted at shorter distances. In principle, one would expect stringy corrections to modify the supergravity answer (mostly)  below $r^*$, so to give back a vanishing $\tau_U$ for $r < r^*$. 

Any other flow not belonging to one of the above classes, is either ending on an unacceptable singularity, or it describes QFTs not admitting a FZ multiplet, for which the conjecture \eqref{buicanconj} is not defined.  We find rather compelling (and a nice AdS/CFT consistency check), that for those flows that are acceptable from a gravitational point of view the monotonicity properties of $\tau_U$ hold, while for all flows for which, for a reason or another, the solution has not to be considered, they would not.

There is a number of interesting directions which may be pursued. For one thing, one could try and consider more complicated models, admitting a larger symmetry group, and see whether qualitatively different results hold. In order to check if our interpretation of the behavior in singular flows is correct, instead, it would be interesting to try and uplift this all discussion to ten dimensions, and look for smooth backgrounds with the desired properties. Finally, it would be very interesting to consider models with emergent symmetries (which our simple supergravity model cannot describe), for which the weakest version of the conjecture, eq.\eqref{buicanconj}, has not been proven in field theory. In fact, holography could in principle provide testable example exactly in the regime which is more difficult to handle with field theory techniques, that is when both the UV and IR fixed points are interacting CFTs. We are not aware of any holographic-flow solution preserving at least four supercharges and enjoying 
emergent abelian symmetries in the IR. Progress in any of these directions would be desirable.
 
%%%%%%%%%%%%%%%%%%%%%%%%%%%%%%%%%%%%%%%%%%%%%%
\section*{Acknowledgements}

We thank Matthew Buican and Gianguido Dall'Agata for correspondence, and Riccardo Rattazzi for discussions. We are grateful to Riccardo Argurio and Diego Redigolo for numerous discussions, and for feedback on a first version of the draft. We acknowledge partial financial support by the MIUR-PRIN contract 2009-KHZKRX.

%%%%%%%%%%%%%%%%%%%%%%%%%%%%%%%%%%%%%%%%%%%%%%%%%%%%%

\bibliographystyle{plainnat}

 \end{document}